# Modeling society with a responsible elite

Tsodikova Ya.Yu., V.A. Trapeznikov Institute of Control Sciences of the Russian Academy of Sciences, Moscow

Chebotarev P.Yu., The Technion, Haifa

**Abstract.** Within the framework of the ViSE (Voting in a Stochastic Environment) model, we examine the dynamics in a society, part of which can be considered an elite. The model allows us to analyze the influence of social attitudes, such as collectivism, individualism, altruism on the well-being of agents. The dynamics is determined by collective decisions and changes in the structure of society, in particular, by the formation of groups of cooperating agents. It is found that the presence of a "responsible elite", combining the support of other agents with limited concern for their own benefit, stabilizes society and eliminates the "pit of losses" paradox. The benefit to society from having a responsible elite is comparable to that from having a prosocial group of the same size. If the elite radically increases the weight of the group component in its combined voting strategy, then its incomes rise sharply, while society's incomes decline. If, in response to the selfish transformation of the elite, a new responsible elite emerges, proportionally larger than the previous one, then society will stabilize again, and the old elite will lose its dominant position. This process can be repeated as long as the size of society allows the formation of new responsible elites of the required size.

# Моделирование общества с ответственной элитой[1]

Я.Ю. Цодикова, Институт проблем управления РАН, Москва

П.Ю. Чеботарев, The Technion, Хайфа

**Аннотация.** В предположениях модели ViSE (голосования в стохастической среде) исследуется социальная динамика в обществе, часть которого может рассматриваться как элита. Модель позволяет анализировать влияние социальных установок, таких как коллективизм, индивидуализм, лоббирование, альтруизм и других, на благосостояние агентов. Динамика благосостояния определяется коллективными решениями и изменением структуры общества, в частности, образованием групп кооперирующихся агентов. Установлено, что наличие "ответственной элиты", комбинирующей поддержку других участников с ограниченной заботой о собственном благе, стабилизирует общество и устраняет парадокс "ямы ущерба". Выгода общества от наличия ответственной элиты сравнима с выгодой от присутствия просоциальной группы того же размера. Если элита радикально увеличивает вес групповой составляющей своей комбинированной стратегии голосования, то ее доходы резко растут, а доходы общества снижаются. Если в ответ на эгоистическую трансформацию элиты возникает новая ответственная элита, пропорционально превышающая предыдущую по численности, то общество вновь стабилизируется, и прежняя элита (ныне "клика") теряет свое доминирующее положение. Этот процесс может повторяться, пока размер общества допускает образование новых ответственных элит требуемой численности.

**Ключевые слова**: *голосование, стохастическая среда, модель ViSE, ответственная элита, просоциальность, кооперация.*

Классификация JEL: D72, D64, C63, D73.

## 1. Введение

### 1.1. Проблема кооперации

Базовая дилемма социальности – выбор между кооперацией и отказом от нее. Эта тема

---

является предметом большого числа работ (см., например, (Axelrod, 2006; Nowak, 2006; Васин, 2010; Gross, et al., 2023)). Указанная проблема наиболее сложна в случае многих агентов, допускающем разные сценарии кооперации. Что эффективнее: небольшое объединение, большой союз или действия в одиночку? Значениями каких параметров внешней среды (благоприятность, стабильность и др.), социальной конфигурации (близость интересов агентов, уже сложившаяся кооперация) и характера взаимодействия определяется ответ на этот вопрос?

В реальности отвечать на него приходится во многих случаях. Главные субъекты международной политики – государства, отстаивающие свои "национальные интересы" (Art, et al., 2023). Эти субъекты объединяются в блоки и союзы, часто противостоящие друг другу. Какие альтернативы этой практике (Cao, 2023) наиболее конструктивны? В разных сегментах рынка, регионах и странах мелкий и средний (не "кооперированный") бизнес чувствует себя хорошо либо проигрывает гигантам. Соотношение преимуществ и издержек нередко можно рассчитать, но в отношении социальных объединений провести такой расчет сложнее. Граждане создают профессиональные, территориальные, политические, правозащитные, религиозные, досуговые и другие объединения. Они могут не тратить на это силы, но потом оказаться неспособными решить возникающие проблемы, беззащитными перед катаклизмами, криминалом или государством. Создаваемые ими ячейки гражданского общества могут укрупняться либо создавать конфедерации (Чеботарев, 2015) для добровольной взаимной поддержки. Это может сочетаться с сотрудничеством с бизнесом и государственными структурами (Полтерович, 2022а). Какие из этих объединений (форм сотрудничества) наиболее эффективны? Какие возможны подходы к исследованию этого точными методами?

Теоретик демократии и прогресса, академик, активный участник Великой французской революции маркиз де Кондорсе писал: "Социальные и природные события в равной степени поддаются счету, и для сведения всего в природе к законам, подобным тем, что открыл Ньютон с помощью математического анализа, все, что нужно, – это достаточное число наблюдений и развитые математические средства" (см. (Шродт, 1997)). Однако для формулирования подобных законов необходим переход от наблюдений к измерениям. У Ньютона, его предшественников и последователей были линейка, весы, часы и иные приборы, и законы устанавливали связи между их показаниями.

Возвращаясь к кооперации, можно утверждать, что она возникает, когда агенты формулируют общие интересы и принимают решения, исходя из них. Но даже когда предмет кооперации – совместная прогулка для удовольствия, если кому-то потребовалась экстренная медпомощь, то другие жертвуют продолжением прогулки ради ее оказания. В целом, как и в этом примере, почти любая кооперация предполагает жертву определенными индивидуальными интересами ради согласованных общих интересов. И именно это создает *проблему взаимной кооперации*: агент, входящий в объединение, с одной стороны, получает поддержку и расширяет возможности, а с другой, как правило, берет на себя обязательства оказывать поддержку партнерам и подчиняться не всегда приятным общим решениям, и это обременение может перевесить получаемые преимущества. Выбирать ли кооперацию в этих условиях, и если да, то как найти ее наилучшую форму?

**1.2. Методологические замечания**

Математический анализ проблемы взаимной кооперации требует введения величин, выражающих удовлетворенность агентов и их возможных групп, а также реализации механизма, преобразующего индивидуальные и кооперативные решения в исходы, успешность которых характеризуется введенными величинами. Поскольку в социальной реальности, как правило, (а) нелегко выделить такие характеристики, измеримые в единой шкале (в частности, из-за субъективности оценок удовлетворенности), и (б) действует множество побочных факторов, маскирующих проявление закономерностей, имеет смысл начинать исследование с упрощенных моделей. Упрощение – один из универсальных подходов к поиску закономерностей; как отмечал Эрроу (см. Рубинштейн, 2013), "Экономика настолько сложна, что без математики, упрощающей реальный мир, понять ее невозможно". К социальной реальности это относится в не меньшей степени. Модель ViSE, исследуемая далее, – одна из таких упрощенных моделей.

Когда анализ модели выявляет закономерности, связывающие ее параметры (отдаленное подобие упомянутых Кондорсе ньютоновских законов), исследователь возвращается к реальности в попытке выявить в ней проявление тех же закономерностей. Если они действительно обнаруживаются, это углубляет понимание социальных процессов и указывает пути направления их в нужное русло. Но программа максимум (нечасто реализуемая) – провести в реальности измерения для идентификации параметров модели, чтобы получить относительно (в силу влияния неучтенных факторов) точные предсказания.

Отметим класс часто используемых для теоретического анализа этих вопросов игровых моделей (самая известная из них – модель дилеммы заключенного), которые фокусируются на последствиях индивидуальных решений о кооперации либо отказе от нее, принимаемых в условиях дефицита информации о решениях других игроков. Эта проблема важна, но она выходит на первый план, когда вопрос о выгодности комбинаций индивидуальных решений уже исследован, иными словами, когда матрицу выигрышей агентов, выбравших свои стратегии, можно считать известной. Тем самым указанная выше проблема взаимной[2] кооперации выносится за скобки, считается заранее решенной либо тривиальной.

В реальности это часто не так, и тогда исследование нужно начать с более раннего этапа – нахождения элементов матрицы выигрышей. В частности, как отмечено выше, взаимный отказ от кооперации может в отдельных случаях быть выгоднее взаимной кооперации. Данными, на основании которых ищется матрица выигрышей, обычно являются: специфика деятельности агентов, механизмы, преобразующие их решения в исходы, и свойства внешней среды, с которой агентам приходится взаимодействовать. В частности, – насколько эта среда благоприятна и предсказуема. И если технические преимущества кооперативной или автономной деятельности связаны с характером этой деятельности, то влияние свойств среды на выгодность кооперации – интересный и нетривиальный вопрос. Из опыта мы знаем, что влияние таких факторов существенно, поскольку в одних средах люди интуитивно выбирают действовать по одиночке, в других создают мини-группы, в третьих – конкурирующие большие, а иногда – общее объединение.

Модель ViSE (Voting in Stochastic Environment), используемая в данной статье, предназначена для анализа эффективности социальных стратегий и процедур принятия решений в зависимости от внешних условий. В ее рамках может быть рассмотрен ряд механизмов кооперации, сводящихся к голосованию за общие (при разных формализациях этого понятия) интересы. Анализ модели помогает выявить некоторые феномены, связанные как с конкретными правилами голосования в стохастической среде, так и с социальностью как таковой (почти всегда предполагающей формирование общих решений или мнений).

Правила "большинства" с порогом (простое, квалифицированное большинство, единогласие, инициативное меньшинство) латентно встроены едва ли не в любую социальную структуру. Это относится и к авторитарным системам, т.к. автократ, не опирающийся на поддержку большинства в определенных группах, рискует потерять власть. В армиях, где действует единоначалие, есть военные советы – в роли консультативных органов. В еще большей мере это относится к структурам гражданского общества. Простое большинство занимает центральное место среди процедур большинства в силу распространенности.

В рамках модели ViSE может быть, в частности, исследован вопрос, насколько кооперация группы агентов выгодна не только для них, но и для остальных. Этот вопрос имеет прямое отношение к практике. Так, наличие в коллективе не очень большой, но сплоченной фракции ("землячества" и т.п.) часто позволяет ей навязывать свою волю остальным в случае их разобщенности.

---

[2] Определение "взаимная" фиксирует отличие от ситуаций, где кооперация – одностороннее решение агента, которое может не подкрепляться решениями предполагаемых партнеров.

### 1.3. Просоциальное поведение

В данной работе, кроме кооперации, мы анализируем также другой тип поведения. Его часто называют альтруистическим, и мы не отказываемся от этого термина, но более точен операциональный (т.е. характеризующий не мотивы,[3] в модели не представленные, а действия) термин "просоциальное поведение" (Thielmann, et al., 2022; Pfattheicher, et al., 2022). От кооперации оно отличается тем, что агенты поддерживают также и тех, кто их не поддерживает ("тех, кто не платит"). Во многих игровых моделях это поведение трактуется как своего рода поражение, когда агент выбирает кооперацию с агентами, сделавшими иной выбор, и, как правило, получает самый низкий выигрыш.

Тем не менее, просоциальное поведение характерно для заметной части людей, и именно эти люди выше всего ценятся многими культурами. Тем самым культуры формируют соответствующие ориентиры, мотивирующие людей (некоторые современные механизмы выработки таких ориентиров описаны в (Полтерович, 2022a,b)). В силу этого просоциальное поведение имеет безусловную ценность, а значит, и *полезность* для самих субъектов. Но если оно реализуется в сфере материально-денежных взаимодействий, то исследователь должен отделять данную полезность от материальной, поскольку эти полезности не могут заместить друг друга ("не хлебом единым", но и "святым духом не насытишься"). Неустранимая потребность просоциального агента в материальном мотивирует рассмотрение так называемых *комбинированных стратегий* голосования, в частности, соответствующих гибриду просоциального и группового (т.е. взаимно кооперативного) поведения, соединенных в определенной пропорции в духе "просвещенного интереса". Последний может формироваться под действием опыта серийного участия во многих взаимодействиях (Axelrod, 2006) и/или восприятия сообщества партнеров как аналог семьи (Васин, 2010). Группа агентов, имеющих такую стратегию, далее будет названа "ответственной элитой". Смысл термина: они – элита, поскольку ответственны. Сущность этой ответственности будет конкретизирована.

Выход за рамки классической рациональности посредством рассмотрения просоциального поведения – один из важнейших трендов в работах по теоретической экономике и социальному выбору (social choice) с начала 1980-х гг. (см. (Margolis, 1982), где альтруизм отождествляется с чувством социальной ответственности). Этот переход обсуждался в ряде дискуссий и в экономической литературе (см., например, (Рубинштейн, 2016; Полтерович, 2016)).

При исследовании просоциального и комбинированного поведения важны, в частности, следующие вопросы.

1. Может ли небольшое число просоциальных агентов оказать серьезное влияние на состояние общества?
2. Тот же вопрос, но применительно к агентам, имеющим комбинированные стратегии. Есть ли пропорция просоциального и группового поведения, которая сохраняет решающее позитивное влияние на состояние общества и обеспечивают агентам с комбинированной стратегией уровень благосостояния не ниже уровня других агентов?
3. Могут ли те же цели быть достигнуты в обществе, где "против" агентов с

---

[3] Именно связь с мотивами, которые сложно дифференцировать не только внешнему наблюдателю, но порой и самому индивиду, вызывает проблемы с термином "альтруизм". Если помощь людям приносит ему удовлетворение, чувство востребованности, самоуважение, ответную теплоту, признание, высокий социальный статус, снижение налогов и т.д., назовем ли его альтруистом? А если она не дает никакого удовлетворения, станет ли он помогать? Является ли основой "истинного" альтруизма лишь эмпатия (Hoffman, 1981) или о нем имеет смысл говорить в терминах мазохизма (Seelig, et al., 2001)? Рассматривая подобные вопросы, некоторые авторы (см., например, (Dibou, 2012; Wilson, 2015)) признают термин "альтруизм" дезориентирующим либо внутренне противоречивым. Понятие просоциального поведения, характеризуя действия, нейтрально к мотивам и потому хорошо вписывается в экономические и политологические модели, где каждый агент имеет целевую функцию, но нет ни мотивов, к которым ее можно было бы редуцировать, ни необходимости в такой редукции.

комбинированной стратегией действует сравнимое число агентов с групповой стратегией?

В статье даются ответы на эти вопросы для некоторых постановок задач. Исследование проводится в рамках модели ViSE.

### 1.4. Модель ViSE

Модель ViSE (Voting in Stochastic Environment) (Борзенко и др., 2006; Чеботарев и др., 2018) – это модель социальной динамики, определяемой коллективными и индивидуальными решениями в стохастической среде. Модель позволяет анализировать влияние процедур принятия решений и стратегий участников (индивидуализм, коллективизм, просоциальность, лоббирование и др.) на благосостояние участников и всего общества.

В модели ViSE каждый *участник* (*агент*) имеет текущий *капитал* (действительное число; для отдельных агентов он может совпадать с *полезностью*), и все $n$ участников голосуют за предложения, генерируемые внешней стохастической средой. Каждое *предложение* представляет собой вектор изменений капиталов агентов; компоненты этого вектора в простейшем случае – реализации независимых одинаково распределенных случайных величин, имеющих математическое ожидание и дисперсию. Последние характеризуют степень благоприятности и разброс предложений внешней среды. Рассматриваются различные классы распределений предложений. Связь модели ViSE с реальностью подробно обсуждалась в (Максимов и др., 2020, раздел 2). Обычно предполагается, что агенты имеют полную информацию о каждом предложении среды.

Начальные условия задаются вектором начальных капиталов агентов. Каждый агент оценивает предложение, используя свой *алгоритм голосования*. Этот алгоритм называется также *стратегией*. Подчеркнем, что в модели ViSE стратегия – это не решение агента, поддерживать ли конкретное предложение, а алгоритм принятия таких решений. В частности, *индивидуалист* (1-*агент*) голосует за предложения, увеличивающие его капитал, члены *группы* (*клики*) поддерживают предложения, увеличивающие совокупный капитал этой группы (*групповая стратегия*), *просоциальные агенты* (*альтруисты*) поддерживают предложения, увеличивающие капитал всего общества (именно такие альтруисты рассматриваются в данной статье) или, например, его беднейшей части. Поскольку капитал общества складывается из капиталов агентов, общественный интерес трактуется здесь в духе методологического индивидуализма. Но модель позволяет рассматривать и общественные интересы, не сводящиеся к индивидуальным.

Принятие или отклонение предложений осуществляется с помощью выбранной *процедуры голосования*. В данной работе это *правило простого большинства*, в ряде других работ – весь класс правил порогового большинства. Принятые предложения реализуются, т.е. капиталы агентов получают приращения, предусмотренные этими предложениями. В случае отклонения предложения капиталы агентов не меняются. Динамика общества определяется изменением капиталов, а также изменением агентами (по тем или иным причинам, предполагаемым программой исследования) своих алгоритмов голосования. Для удобства описания при изменении алгоритмов голосования будем говорить о новом обществе.

Модель изучается в двух вариантах: "с вымиранием", где агент с отрицательным капиталом выбывает либо теряет право голоса, и "без вымирания", где он не поражается в правах. В первом случае наблюдается нетривиальная динамика изменения числа участников. В данной статье исследуется второй, более простой вариант. Здесь динамика сводится к линейной экстраполяции одношаговых результатов, а начальные капиталы играют малосущественную роль одноразовых бонусов. Отметим, что второй вариант может трактоваться как частный случай первого, где величины начальных капиталов практически исключают банкротство за рассматриваемое число шагов.

Таким образом, модель ViSE имеет ряд параметров, варьируя которые можно получать упрощенные образы некоторых реальных процессов. В рамках модели можно искать оптимальные процедуры голосования для обществ разной структуры, принимающих решения в разных средах. Кроме того, она позволяет находить наиболее выгодные для агентов и для общества алгоритмы голосования; оценивать последствия формирования групп и иных фракций. В рамках модели можно также сравнивать схемы стимулирования (путем перераспределения доходов) участников, чьи социальные установки помогают обществу (Афонькин, 2021). Как было отмечено, ценность подобного моделирования в том, что закономерности, выявленные с помощью простых моделей, часто проявляются и в реальных процессах, где без упрощения их сложно обнаружить в силу субъективности оценок и действия множества маскирующих факторов. Выявленная же закономерность может быть учтена и использована.

Модель ViSE изучалась в ряде работ, в частности, (Борзенко и др., 2006; Чеботарев, 2006, 2021; Афонькин, 2021; Чеботарев и др., 2008, 2016, 2018; Malyshev, 2021; Максимов и др., 2020). Одним из ее отличий от других сопоставимых моделей (Penn, 2009; Dziuda, Loeper, 2015, 2016; Binmore, Eguia, 2017) является то, что фокус исследования переносится с поиска равновесия в переговорах участников к максимизации капитала (полезности), иными словами, к выбору эффективных алгоритмов голосования и процедур принятия коллективных решений. В (Hortala-Vallve, 2012), в отличие от процессов в модели ViSE, не подразумевается динамика, и ряд предложений, сформированных экзогенно, равноправно конкурируют, но, как и в модели ViSE, принятие предложения изменяет капиталы/полезности агентов, и по этим изменениям оценивается эффективность правил принятия решений. Подробнее о сравнении модели ViSE с другими моделями см. в (Чеботарев и др., 2018; Максимов и др., 2020; Malyshev, 2021).

### 1.5. Комбинированные алгоритмы голосования

В данной работе исследуются *комбинированные стратегии голосования*. Необходимость в них обусловлена следующими отмечавшимися ранее обстоятельствами.

1. Для общества, состоящего из индивидуалистов, характерно разорение агентов в неблагоприятной среде в результате реализации решений большинства (так называемый парадокс "ямы ущерба" (Чеботарев и др., 2016; Malyshev, 2021)).
2. Наличие сравнительно небольшой фракции альтруистов защищает других агентов (и общество в целом) от разорения, однако сами просоциальные агенты оказываются малообеспеченным слоем, подверженным разорению.

Чтобы "спасти спасателей", можно использовать два подхода: оплатить их усилия, собрав налоги со спасенных (Афонькин, 2021) (что может быть моделью идеальной бюрократии), либо позволить им "спасать" себя несколько интенсивнее, чем остальных, что выглядит своеобразной, как сказали бы в XVIII веке, "акциденцией от дел". В наши дни продвижение решений в интересах влияющей на них группы описывается понятиями политического дохода / бюрократической ренты, а у широкой публики ассоциируется с коррупцией или лоббированием. Всегда ли это "подтачивает" общество? А если практикуется в малых или "гомеопатических" дозах? Это один из вопросов, исследуемых в данной статье.

Речь идет об альтруизме, к которому добавляется ограниченная мера защиты *групповых* интересов, что реализуется посредством *комбинированных стратегий голосования*. А именно, агент оценивает предложения, находя выпуклую комбинацию среднего прироста капитала всего общества и среднего прироста капитала группы, членом которой является. Если эта комбинация положительна, он поддерживает предложение, в противном случае голосуют против него.

Данная целевая функция, как и "общая полезность" (total utility) Марголиса [Margolis, 1982, pp. 11, 17], в целом согласуется с концепцией Бергсона-Самуэльсона (Samuelson, 1977) и может быть использована для описания варианта коллаборативизма (Полтерович, 2016). В качестве другого примера отметим "целевые функции производителей[, которые] состоят из двух слагаемых: полезности от их прибыли (эгоистическая компонента) и полезности от совокупной прибыли всех участников, взятой с некоторым коэффициентом (альтруистическая компонента)" (Полтерович, 2022b). Подчеркнем, что в нашем случае речь идет о коллаборативизме, комбинирующем общественный интерес не с личным, а с групповым. О некоторых других реализациях концепции Бергсона-Самуэльсона), состоящей в агрегировании общественных и индивидуальных ценностей, см. Рубинштейн, 2016.

Фракцию, члены которой придерживаются указанной комбинированной стратегии голосования при достаточно высокой доле ее просоциальной составляющей, можно рассматривать как "ответственную элиту". В следующих разделах мы обсудим возможную эволюцию обществ, в которых появляется такая фракция.

## 2. Общество с просоциальными агентами

### 2.1. Может ли небольшое количество альтруистов предотвратить разорение общества?

Рассмотрим общество, состоящее из 101 агента. В качестве процедуры голосования примем правило простого большинства голосов всех агентов.

Число агентов выбрано нечетным для достижения большей лабильности: предложения могут приниматься перевесом всего в один голос; исключено равенство голосов за и против. Число участников предположительно достаточно велико, чтобы могли проявиться интересующие нас качественные закономерности. Для их выявления рассмотрим общество в динамике: некоторые агенты будут менять свои алгоритмы голосования, руководствуясь идеями рациональности, трактуемой в широком смысле.

"Финансовая" успешность участника будет характеризоваться математическим ожиданием изменения его капитала за один раунд (шаг) голосования. Назовем эту величину *средним приростом капитала* (*СПК*) участника. Пусть вначале общество состоит из 1-агентов: каждый голосует за предложение тогда и только тогда, когда оно увеличивает его капитал.

Поскольку при отклонении предложения капиталы участников не меняются, СПК агента равен произведению ожидаемого прироста его капитала при условии принятия предложения на вероятность такого принятия. СПК зависит от числа агентов, их алгоритмов голосования, процедуры агрегирования голосов и распределения, генерирующего предложения (его формы, среднего и дисперсии, которые предполагаются конечными). При неизменных алгоритмах голосования процесс одобрения предложений вероятностно стационарен, и мы будет характеризовать его значениями СПК агентов.

В качестве генераторов предложений рассмотрим нормальные распределения $N(\mu, 80)$. Величина $\sigma = 80$ выбрана для определенности; диаграммы в координатах $\mu/\sigma$, СПК$/\sigma$ инвариантны к $\sigma$: $\sigma$ играет роль масштаба. В работах (Чеботарев и др., 2018; Максимов и др., 2020) исследовалась специфика генераторов с длинными/тяжелыми хвостами.

На рис. 1 показана зависимость СПК от математического ожидания $\mu$ при указанных предположениях. Эта функция найдена аналитически (Чеботарев, 2006; Чеботарев и др., 2016); последующие кривые в случаях, когда результаты интегрирования не выражаются в стандартных функциях, получены численным моделированием.

Рис. 1. Средний прирост капитала (СПК) агента за один раунд голосования

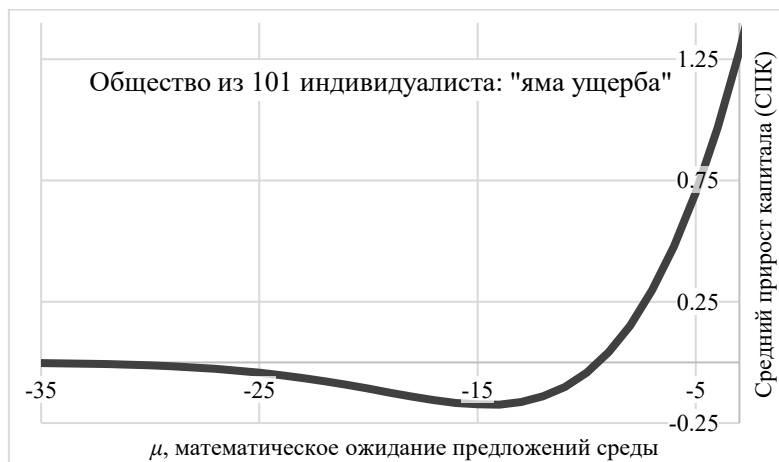

101 индивидуалист, распределения $N(\mu, 80)$, простое большинство.

При $\mu \in [-30, -10]$ СПК агента принимает заметные отрицательные значения, т.е. общество неизбежно беднеет вследствие реализации решений большинства; кривая СПК имеет так называемую *яму ущерба*. Тем самым в предположениях модели ViSE нарушается принцип Адама Смита (в интерпретации (Рубинштейн, 2016)) "Если каждый будет стремиться к своей корысти, то "невидимая рука" провидения приведет к всеобщему благосостоянию". Механизм парадокса "ямы ущерба" состоит в том, что большинство, одобряющее предложение, получает в существенно неблагоприятной среде в среднем меньше, чем теряет дополняющее его меньшинство. Этот механизм, в частности, показывает, что парадокс тотально-мажоритарного пути (см. (Миркин, 1974, п. 2.1.3)), известный также как парадокс А.В. Малишевского, реализуем без намеренного манипулирования повесткой дня, т.е. имеет более глубокую природу.

Парадокс "ямы ущерба" можно преодолеть, увеличив порог голосования (Чеботарев и др., 2016; Malyshev, 2021) либо изменив алгоритмы голосования агентов. В данной работе мы рассмотрим второй путь, зафиксировав в качестве процедуры голосования правило простого большинства, имеющее самый широкий круг применений.

Предположим теперь, что в обществе появились агенты *с просоциальной стратегией*: эти агенты (которых для краткости иногда называем *альтруистами*) голосуют за те и только те предложения, которые увеличивают совокупный капитал общества.

На рис. 2 показаны кривые зависимости СПК в обществе от числа $k$ агентов с просоциальной стратегией, если остальные участники – индивидуалисты. Приведены графики для случая $\mu = -14$, в котором СПК агентов в обществе индивидуалистов близок к минимуму (рис. 1). Под "СПК общества" здесь и далее понимается СПК случайно выбранного агента, или, иными словами, средний СПК всех агентов. Для сравнения на рис. 2 показан СПК общества из 101 1-агента, обозначенный 101E.

Рис. 2. СПК при росте числа агентов с просоциальной стратегией.

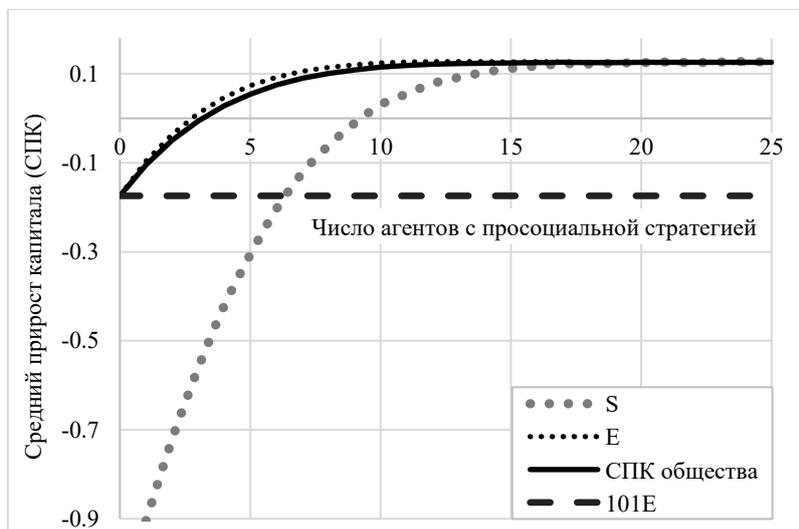

СПК агента с просоциальной стратегией (S), 1-агента (E) и среднее значение по обществу в зависимости от числа просоциальных агентов: 101 агент, генератор предложений $N(-14, 80)$.

При увеличении $k$ влияние группы просоциальных агентов на решения, принимаемые большинством, усиливается, что приводит к росту СПК всех участников.

Отметим следующие закономерности этого роста.
1. При $k \geq 3$ СПК индивидуалистов положителен – в среднем они выигрывают от принимаемых решений.
2. При $k \geq 4$ общество в целом уже не разоряется.
3. При $k \geq 7$ СПК альтруистов выше, чем СПК в обществе, состоящем из индивидуалистов.

4. При $k \geq 10$ СПК альтруистов положителен: они в среднем выигрывают от принимаемых решений.
5. При $15 < k \leq 18$ 1-агенты лишь незначительно опережают альтруистов; при $k > 18$ их преимущество нивелируется.

Таким образом, общество, разорявшееся (в силу парадокса ямы ущерба) из-за решений большинства, перестает разоряться уже при 4 альтруистах. Если альтруистов не меньше 10, то и они в среднем уже не беднеют, а если больше 18, то их СПК не отстает от этого показателя других участников. При росте $n$ необходимая для этого доля просоциальных агентов в обществе снижается пропорционально $1/\sqrt{n}$. Численное выражение этих закономерностей определяется статистической природой предложений, но качественно они объясняются ростом доли минимальных решающих коалиций, включающих альтруистов, с ростом числа альтруистов.

### 2.2. Пороговое сравнение успешности участников

Общества с просоциальными участниками рассматривались в (Tsodikova, 2020), где СПК агентов с разными стратегиями сравнивались при значении $\mu$, минимизирующем СПК в обществе из индивидуалистов, т.е. в худшем случае (на "дне ямы ущерба"). Однако значение $\mu$, минимизирующее СПК, зависит от структуры общества, поэтому более информативно сравнение при *всех* отрицательных значениях $\mu$. Этот более универсальный подход реализован в данной работе[4]. Однако при его использовании в некоторых случаях утрачивается однозначность результата сравнения категорий участников. Например, при низких значениях $\mu$ СПК шести альтруистов превосходит СПК 101 1-агента, а при более высоких $\mu$ имеем противоположный результат сравнения (рис. 3).

Рис. 3. Неоднозначность сравнения СПК двух категорий агентов.

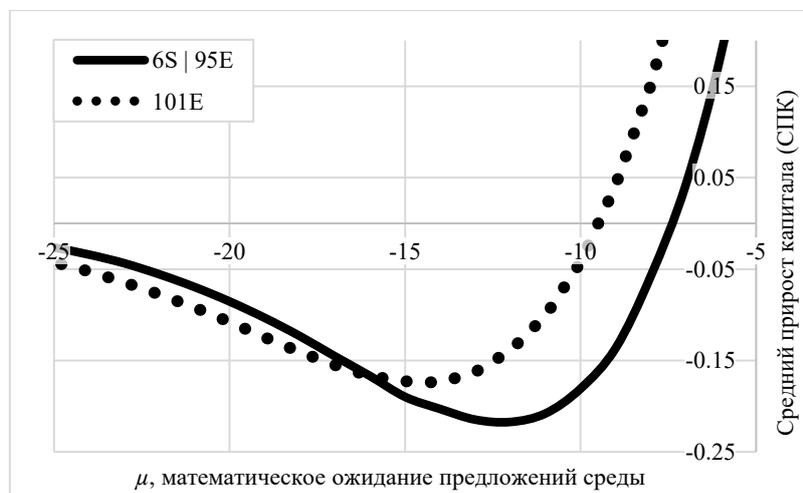

Сравнение СПК 6 агентов с просоциальной стратегией в обществе с 95 индивидуалистами и СПК 101 1-агента.

Чтобы снизить число случаев неоднозначности посредством пренебрежения малым превосходством, введем порог неразличения: различие меньше этого порога будет рассматриваться как равенство. В качестве порога взята одна сотая модуля наименьшего по $\mu$ значения СПК в обществе из 101 индивидуалиста при генераторе предложений $N(\mu, 80)$. Эта величина округляется до 0,0017. Если не оговорено иное, "больше/меньше в заданной точке" означает "больше/меньше в этой точке на величину, превосходящую порог". При сравнении СПК на всей оси $\mu$ "больше" будет означать "больше хотя бы в одной точке и ни в одной не меньше".

---
[4] Аналог ямы ущерба в благоприятной среде рассмотрен в (Afonkin, et al., 2023).

### 2.3. Переход от просоциальной стратегии к комбинированной

Как установлено в разделе 2.1, в модели ViSE небольшая фракция просоциальных агентов может защитить общество от разорения. Поэтому такие агенты очень ценны для общества.

В то же время СПК просоциальных агентов не выше СПК индивидуалистов (а часто значительно ниже). Та же закономерность типична для реальных обществ, поэтому просоциальных агентов в них не так много. В неблагоприятной среде такие агенты, спасая общество, сами могут разориться, что иллюстрирует рис. 2. В связи с этим представляется важной задача выработки механизмов материальной поддержки просоциального поведения.

Одним из подходов к ее решению, как отмечено выше, является перераспределение доходов участников в пользу просоциальных агентов с помощью налоговых схем (Афонькин, 2021). Такие схемы требуют централизованных механизмов реализации. Использование описанных в разделе 1.5 комбинированных целевых функций – подход на уровне индивидуальных стратегий агентов.

Далее рассмотрим комбинированные стратегии голосования, совмещающие помощь всему обществу и поддержку тех, кто ее оказывает.

## 3. Динамика обществ с "элитой"

### 3.1. Фракция с комбинированной стратегией как ответственная элита

Пусть в составе общества имеется фракция агентов со следующей комбинированной стратегией голосования. Оценивая предложение, агент находит величину $(1-\alpha)D_1 + \alpha D_2$, где $D_1$ и $D_2$ – средние приращения капитала соответственно всех $n$ агентов и агентов фракции, к которой принадлежит данный агент, согласно рассматриваемому предложению; $\alpha \in (0, 1)$ – параметр, задающий степень группового эгоизма данной фракции. Агент поддерживает предложение, если $(1-\alpha)D_1 + \alpha D_2 > 0$. Таким образом, алгоритм голосования агента использует выпуклую комбинацию просоциальной и групповой целевых функций.

Если 8 агентов из 101 образуют фракцию с такой стратегией при $\alpha = 0{,}066$, а остальные 93 агента — индивидуалисты, то для обозначения общества используется запись 101:{8(0,934S+0,066G); 93E}, где S, G и E кодируют соответственно просоциальную, групповую и индивидуалистическую целевые функции.

Найдем численность фракции и коэффициент $\alpha$ комбинированного условия, при которых, независимо от $\mu$, (А) средние приросты капитала всех агентов неотрицательны в терминах порогового сравнения и (B) СПК члена фракции с комбинированным условием голосования выше, чем СПК 1-агента (индивидуалиста) в том же обществе.

Рис. 4. Ответственная элита из 8 агентов в обществе 101:{8(0,934S+0,066G); 93E}.

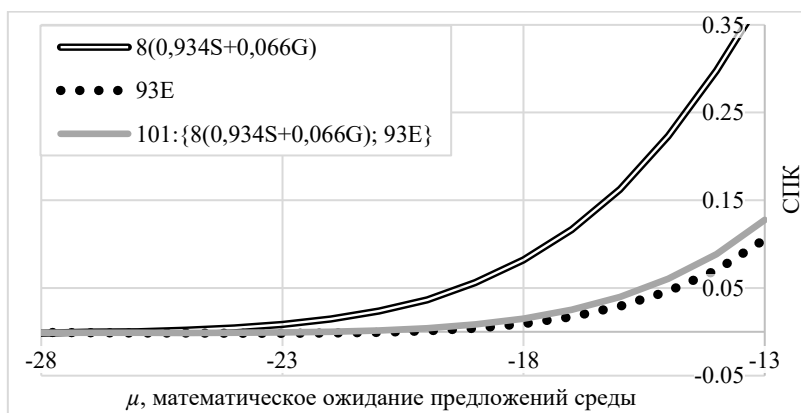

Фракцию, реализующую условия (А) и (В), будем называть *ответственной элитой*.

В рассмотренном примере 8 агентов придерживаются комбинированной стратегии, и их фракция является ответственной элитой при $0{,}055 \le \alpha \le 0{,}066$. Если $\alpha \ge 0{,}067$, то нарушается условие (A), при $\alpha \le 0{,}054$ не выполняется условие (B). При меньшем размере фракции условия (A) и (B) несовместимы.

Пусть для ответственной элиты из 8 агентов $\alpha = 0{,}066$ (рис. 4).

Если средние доходы 8 агентов с просоциальной стратегией при $\mu = -14$ отрицательны и значительно ниже, чем у 1-агентов (рис. 2), то рассмотренная ответственная элита, в силу условий (B) и (A), лидирует в обществе и защищает его от разорения (рис. 4). При этом *средние* доходы в этих двух обществах отличаются мало (рис. 5) и – существенно выше, чем в обществе {101E}.

Таким образом, наличие сравнительно небольшой фракции с комбинированной стратегией голосования позволяет решить две задачи: (A) защитить общество от разорения и (B) обеспечить членам этой фракции средний доход, несколько превышающий средний доход 1-агентов в том же обществе (рис. 4).

Рис. 5. Сравнение СПК обществ с комбинированной/просоциальной стратегией 8 агентов и общества из 1-агентов.

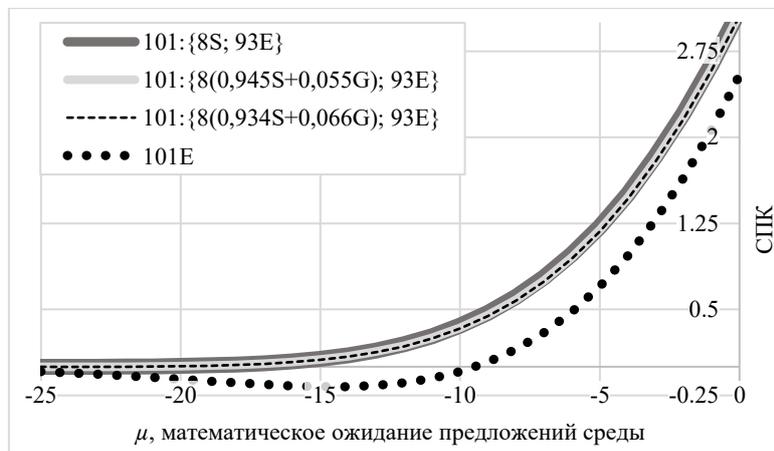

### 3.2. Переход элиты к групповому эгоизму и реакция общества на него

"Ответственную элиту", защищающую общество от разорения и поддерживающую доходы своих членов на уровне, немного превышающем средний по обществу, можно рассматривать как одну из возможных моделей адекватной бюрократии.

Рис. 6. Превращение ответственной элиты в клику.

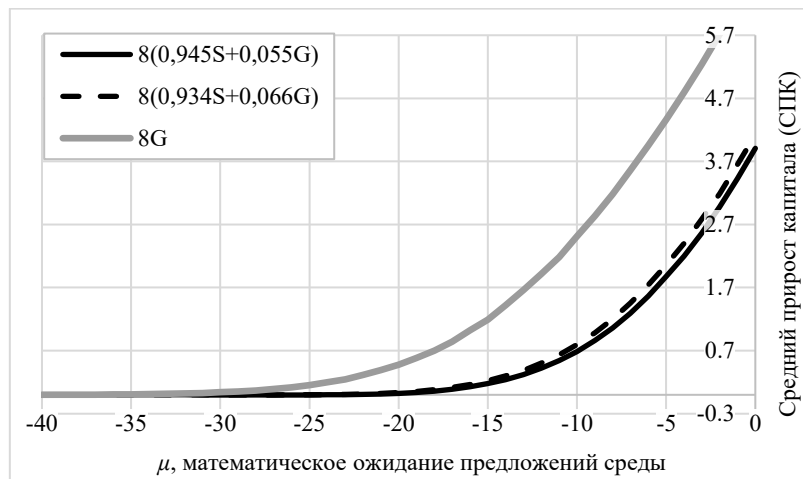

СПК фракции из 8 агентов в обществах 101:{8((1−α)S+αG); 93E} при $\alpha \in \{0{,}055;\ 0{,}066;\ 1\}$.

Подобный слой общества обычно испытывает соблазн повысить свой доход за счет (в терминах данной модели) увеличения веса групповой целевой функции в своей комбинированной стратегии. Иными словами, он может изменить пропорцию: в большей степени влиять на принимаемые решения в собственных интересах и в меньшей – в интересах общества. Кривая, соответствующая весу $\alpha = 1$, показана на рис. 6. Она демонстрирует, что превращение ответственной элиты в клику существенно обогащает ее членов.

В то же время для остальной части общества это изменение, которое можно назвать переходом элиты к групповому эгоизму и пренебрежением социальной ответственностью, крайне невыгодно. Рис. 7 показывает, что в этом случае средний доход агента и доход 1-агента значительно ниже доходов в индивидуалистическом обществе. Фракция, имевшая комбинированную стратегию голосования, становится кликой, которую можно назвать *безответственной элитой.*

Но на ту же ситуацию можно посмотреть и иначе. Клика – это часть общества, выбравшая кооперацию. Кооперация очень выгодна ее участникам, но невыгодна остальным.

Рис. 7. Влияние превращения ответственной элиты в клику на средний прирост капитала членов общества и 1-агентов.

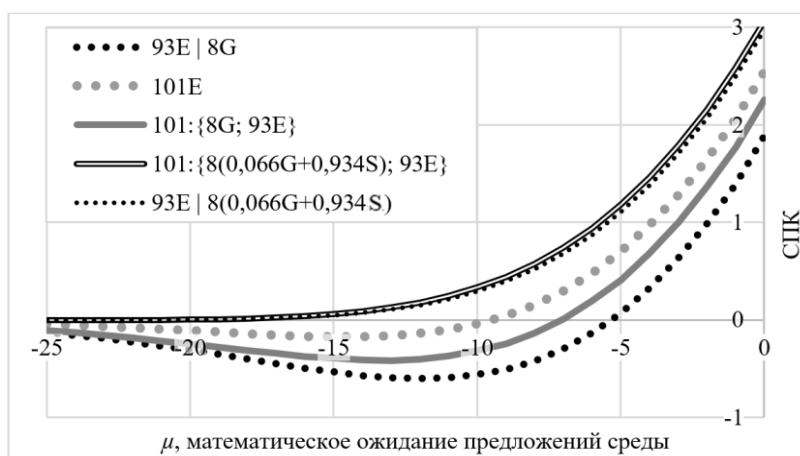

Это стимулирует остальных присоединяться к кооперирующейся группе, увеличивая "снежный ком кооперации" (Чеботарев и др., 2009), в результате чего групповой эгоизм приближается к альтруизму, а СПК всего общества постепенно растет.

Таким образом, механизм "снежного кома кооперации" защищает общество от разорения, но лишь в конечном итоге, а трансформация ответственной элиты в клику сразу же обрекает на разорение не входящих в нее агентов.

Перечислим доступные им защитные меры:

(1) присоединяться к клике ("снежный ком кооперации", что, однако, самой клике, начиная с определенного ее размера, невыгодно (Чеботарев и др., 2009));

(2) сформировать еще одну фракцию с групповой стратегией (Чеботарев и др., 2008);

(3) образовать фракцию с просоциальной или комбинированной стратегией.

Рассмотрим путь (3) – формирование новой фракции с комбинированной стратегией. У этой фракции – две цели, указанные выше: защитить от разорения остальное общество и – самой достичь благосостояния, несколько превышающего среднее. Два ее параметра: число участников и коэффициент комбинированного условия $\alpha$. Выбор их должен обеспечивать выполнение комбинации условия

(А) *СПК агентов всех категорий неотрицателен*

с одним из условий:

(В) *СПК члена новой фракции выше СПК 1-агента*

или

(В+) *СПК члена новой фракции выше среднего по обществу СПК*

или

(В++) *СПК члена новой фракции выше СПК членов всех клик.*

Наименьшая фракция, для которой при наличии группы $8G$ реализуема комбинация (AB++), состоит из 15 агентов. Условие (A) выполняется для нее при $\alpha < 0{,}086$, (B) – при $\alpha \geq 0{,}028$, (B++) – при $\alpha \geq 0{,}063$. Следовательно, общества с минимальным $\alpha$, при котором выполняются комбинации (AB) и (AB++), имеют соответственно вид $\{8G_1; 15(0{,}972S+0{,}028G_2); 78E\}$ и $\{8G_1; 15(0{,}93S+0{,}063G_2); 78E\}$. (AB) реализуемо уже при 13 агентах во фракции $G_2$ (нижним индексом обозначаем номер фракции), поскольку для этого достаточно меньшего ее влияния на принимаемые решения, чем для (AB++).

На рис. 8 показана зависимость пороговых значений $\alpha_A$, $\alpha_B$, $\alpha_{B+}$ и $\alpha_{B++}$ параметра $\alpha$ комбинированной стратегии фракции $G_2$ от ее численности $k = g_2$. При $\alpha < \alpha_A$ выполняется условие (A), при $\alpha > \alpha_B$, $\alpha > \alpha_{B+}$ и $\alpha > \alpha_{B++}$ – соответственно условия (B), (B+) и (B++).

Рис. 8. Пороговые значения условий (A), (B), (B+) и (B++).

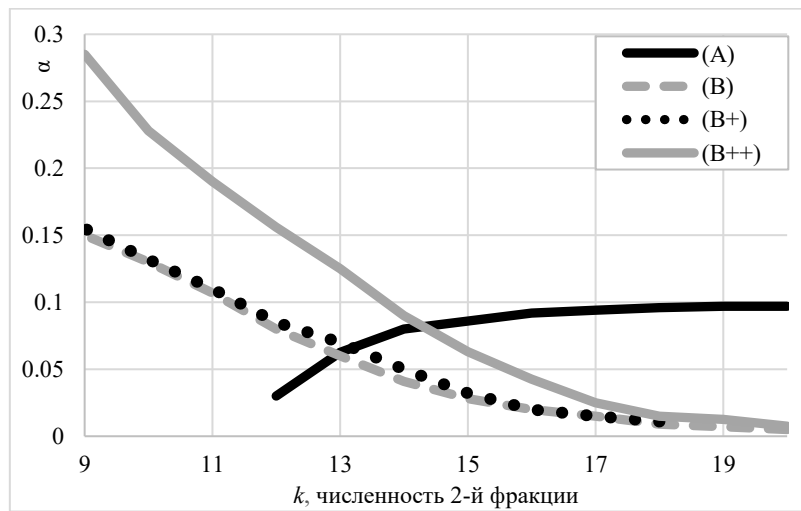

Общества 101:$\{8G_1; k((1-\alpha)S+\alpha G_2); (93-k)E\}$. При $\alpha \leq \alpha_A$ СПК 1-агентов и членов клик неотрицательны; при $\alpha \geq \alpha_B$, $\alpha \geq \alpha_{B+}$ и $\alpha \geq \alpha_{B++}$ СПК фракции $G_2$ соответственно выше СПК 1-агентов, среднего по обществу и членов клик.

### 3.3. Соблазн использования влияния на благо элиты и соответствующая ему социальная динамика

Максимальный доход агенты фракции $G_2$ получают при $\alpha = 1$, т.е. когда они, игнорируя интересы других агентов, голосуют в соответствии с групповой стратегией.

Рис. 9. Вторая фракция превращается в клику: 101:$\{8G_1; 15G_2; 78E\}$. Остальные агенты разоряются.

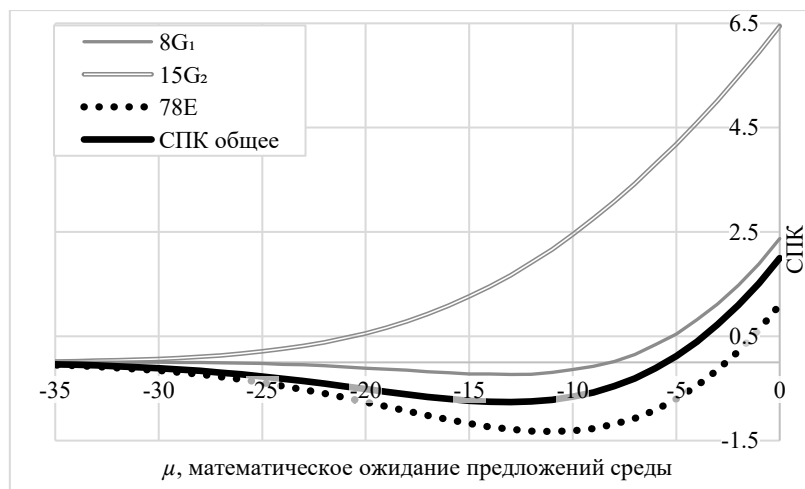

Если фракция $G_2$, как и $G_1$, перешла к групповому эгоизму, то участники других категорий, а также общество в целом разоряются при значениях $\mu$, принадлежащих определенным интервалам (рис. 9).

Это требует от общества очередных защитных мер. Продолжая исследование возможностей фракций с комбинированной стратегией, предположим, что из 1-агентов формируется новая фракция этого типа. Зависимость СПК участников от ее размера при $\mu = -14$ и коэффициенте $\alpha = 0{,}05$ ее комбинированной стратегии показана на рис. 10. Фрагмент для $17 \leq g_3 \leq 30$ при $\alpha = 0{,}045$ и $\alpha = 0{,}08$ представлен на рис. 11, где с ростом $g_3$ ответственная элита обгоняет клику (сохраняющую преимущество перед 1-агентами) тем раньше, чем выше $\alpha$.

Рис. 10. Образование ответственной элиты $G_3$.

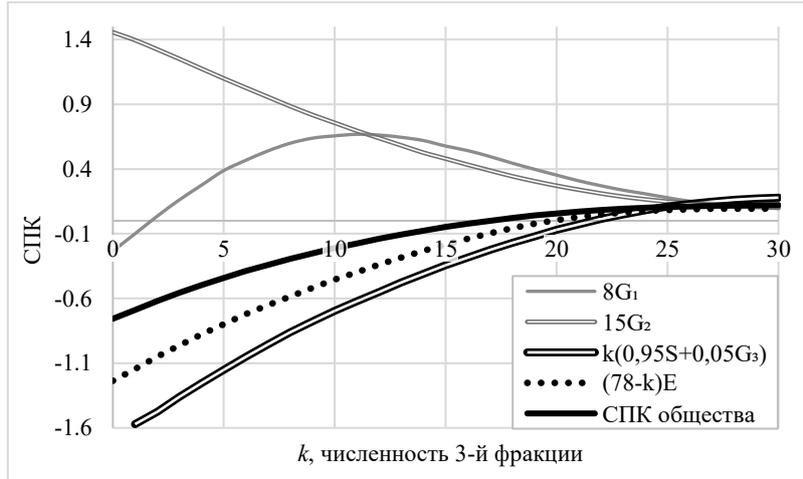

СПК агентов в обществах $101:\{8G_1;\ 15G_2;\ k(0{,}95S+0{,}05G_3);\ (78-k)E\}$ в зависимости от размера $k = g_3$ фракции $G_3$ при $\mu = -14$.

Рис. 11. Образование ответственной элиты $G_3$ с $\alpha = 0{,}045$ или $\alpha = 0{,}08$ при $\mu = -14$.

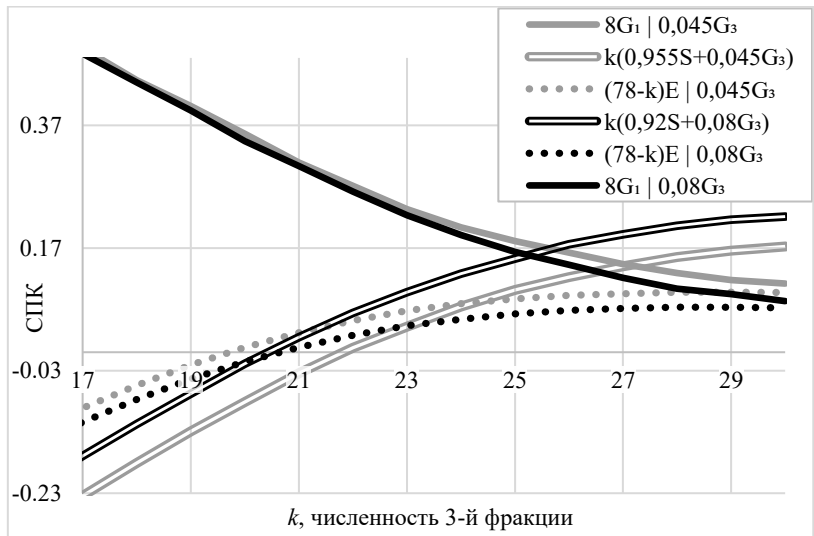

СПК в обществах $101:\{8G_1;\ 15G_2;\ k(0{,}955S+0{,}045G_3);\ (78-k)E\}$ и $101:\{8G_1;\ 15G_2;\ k(0{,}92S+0{,}08G_3);\ (78-k)E\}$ в зависимости от размера $k = g_3$ фракции $G_3$.

Для значений $\alpha = 0{,}045$ и $\alpha = 0{,}08$ параметра комбинированной стратегии фракции $G_3$ найдем ее размер, гарантирующий выполнение условий (А), (В) и (В++) в терминах порогового сравнения при всех значениях $\mu$.

Заметим, что при $\alpha = 0{,}045$ для выполнения условия (А) достаточно $g_3 = 20$; для (В) нужно как минимум 25 агентов, для (В++) – 28 агентов. При $\alpha = 0{,}08$ соответствующие числа: 21, 21 и 26 агентов.

Ранее при рассмотренных превращениях ответственной элиты в клику новая ответственная элита, обеспечивающая выполнение комбинации (AB++), в 1,8–1,9 раза превосходила по численности предыдущую. При имеющихся кликах в 8, 15 и 27 участников оставшихся агентов достаточно для образования фракции, сохраняющей ту же пропорцию.

Как и на предыдущем шаге, после превращения фракции $G_3$ в клику лишь она не разоряется в новом обществе с тремя кликами и индивидуалистами; общество же в целом разоряется (рис. 12). При этом СПК в клике с 15 агентами ниже, чем в клике с 8 агентами.

Рис. 12. Общество с тремя кликами: разоряются общество в целом и все, кроме наибольшей клики.

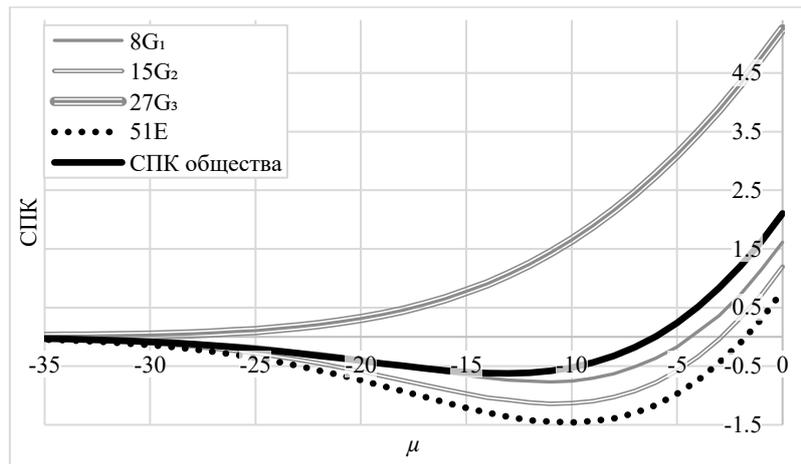

СПК агентов в обществе 101:{$8G_1$; $15G_2$; $27G_3$; $51E$} при разных $\mu$.

Продолжая исследование ответственной элиты, рассмотрим появление и увеличение фракции $G_4$ со стратегией $0{,}9S+0{,}1G_4$. Тогда СПК участников меняется как показано на рис. 13. СПК всего общества положителен при фракции $G_4$ из 31 или более агентов.

Рис. 13. Образование ответственной элиты $G_4$ при $\mu = -14$.

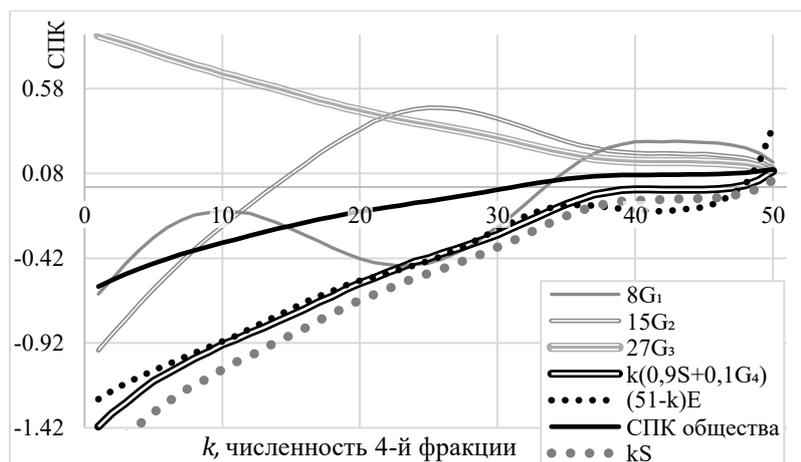

СПК всех агентов в обществах 101:{$8G_1$; $15G_2$; $27G_3$; $k(0{,}9S+0{,}1G_4)$; $(51-k)E$} положителен при $k = 50$. При $k = 51$ $G_4$ – "диктатор".

Колебания СПК разных категорий участников при росте фракции $G_4$ связаны с появлением, исчезновением или изменением вероятности образования различных минимальных решающих коалиций. Заметим, что из шести возможных строгих упорядочений СПК фракций $G_1$, $G_2$ и $G_3$ здесь при разных $\mu$ реализуются пять.

Общая закономерность состоит в том, что если две клики имеют одинаковое влияние на принимаемые решения, т.е. входят в одинаковые (по составу остальных участников) минимальные решающие коалиции, то СПК выше в меньшей клике. Это объясняется законом больших чисел: среднее независимых одинаково распределенных случайных величин с отрицательным математическим ожиданием имеет тем меньшую вероятность положительности и тем меньшее ожидаемое значение при условии положительности, чем больше величин усредняется. Аналогичная закономерность в реальной жизни: чем больше людей, тем меньше вероятность, что им в среднем сильно повезет, и ниже средняя степень этого большого везения.

Если фракция $G_4$ имеет просоциальную стратегию ($\alpha = 0$), то существенное отличие результирующей диаграммы от диаграммы, показанной на рис. 13, – в кривой этой фракции. Эта кривая добавлена на рис. 13 (она обозначена $kS$): как обычно, агенты с просоциальной стратегией имеют более низкий СПК, чем агенты других категорий, и комбинированная стратегия используется для увеличения их благосостояния.

В случае присоединения всех оставшихся 1-агентов к четвертой фракции со стратегией $0{,}9S+0{,}1G_4$ она, с численностью 51, становится "диктатором": общество принимает те и только те предложения, которые она поддерживает. Фрагмент рассмотренной диаграммы, где размер $g_4$ фракции $G_4$ меняется от 44 до 51, показан на рис. 14. На этом интервале СПК фракций $G_1$, $G_2$ и $G_3$ постепенно, но с ускорением снижаются с ростом $g_4$, $G_4$ увеличивает СПК, больше всех выигрывают 1-агенты (преимущество "малой партии" – small party bias (Browne, et al., 1973; Warwick, et al., 2001)), и медленно растет СПК всего общества.

Рис. 14. Фрагмент диаграммы на рис. 13 при $k = g_4 \in [44, 51]$.

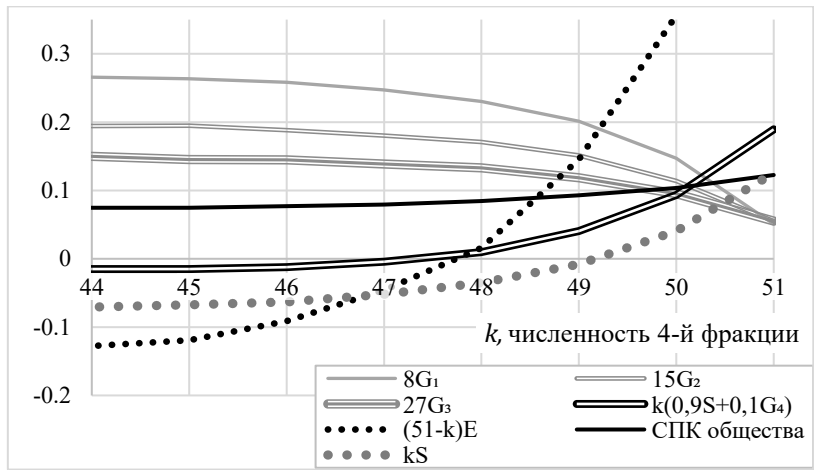

Рис. 15. Фракция $G_4$ удовлетворяет условию (AB++), определяющему ответственную элиту.

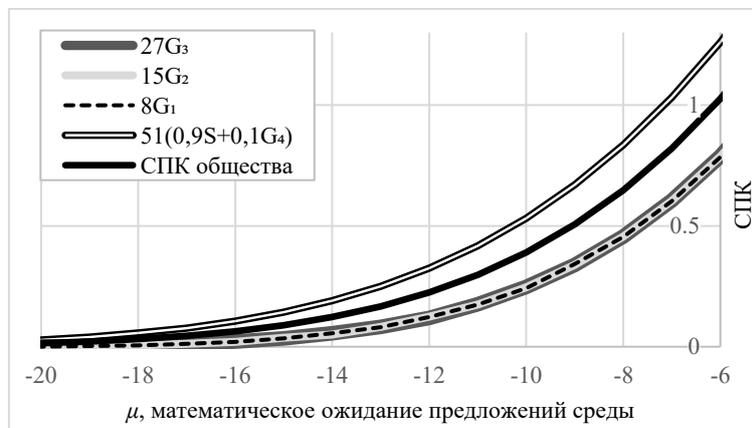

СПК участников в обществе 101:{$8G_1$; $15G_2$; $27G_3$; $51(0{,}9S+0{,}1G_4)$}.

Фракции $G_4$ с $g_4 = 51$, чтобы быть ответственной элитой при всех $\mu$, достаточно придерживаться рассмотренной стратегии $0{,}9S+0{,}1G_4$ (рис. 15). При бóльших $\alpha$ нарушается условие положительности (А).

Если фракция $G_4$ с 51 участником придерживается просоциальной стратегии, то все агенты в обществе имеют равный СПК, увеличивающийся при росте $\mu$ (рис. 16).

Рис. 16. "Диктат" просоциальной фракции, составляющей большинство: СПК всех агентов равны.

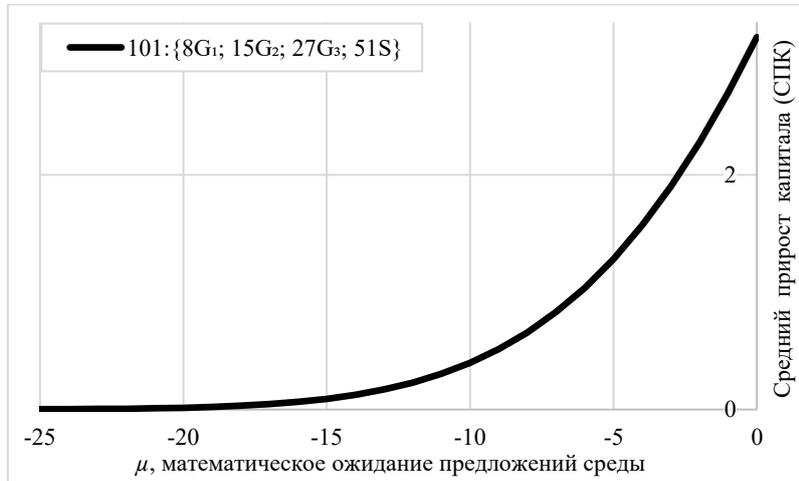

СПК участников в обществе 101:{$8G_1$; $15G_2$; $27G_3$; $51S$} при разных $\mu$.

Специальный интерес представляет общество, в котором фракция $G_4$ состоит из 50 участников, и один агент остается индивидуалистом. Этот агент обладает значительным личным влиянием, определяя принятие/отклонение предложений, по которым расходятся мнения фракции $G_4$ с одной стороны и фракций $G_1$–$G_3$ с другой; в силу этого он лидирует по СПК во всем обществе[5] (рис. 17, где $G_4$ размера 50 придерживается стратегии $0{,}9S+0{,}1G_4$).

Рис. 17. Единственный 1-агент демонстрирует эффект ключевой "малой партии".

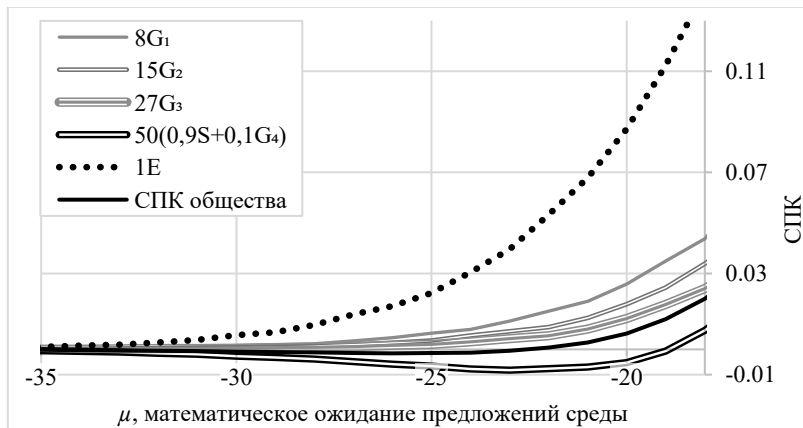

В обществе 101:{$8G_1$; $15G_2$; $27G_3$; $50(0{,}9S+0{,}1G_4)$} 1-агент – ключевой
в двух решающих коалициях, и он вне конкуренции по СПК.

За 1-агентом в порядке убывания СПК следуют члены клик $G_1$, $G_2$, $G_3$, случайно выбранный участник (СПК общества положителен при сравнении с порогом) и агент из фракции с комбинированной стратегией. Такой порядок СПК клик – проявление отмеченного выше следствия закона больших чисел.

Если фракция $G_4$ с 51 участником становится кликой, то она, разумеется, остается диктатором. Тогда СПК ее членов неотрицателен, но СПК всех остальных участников отрицателен при всех $\mu < 0$. СПК случайно выбранного участника отрицателен при пороговом сравнении, когда $-42 \leq \mu \leq -7$ (рис. 18).

---

[5] Этот эффект также заметен, если в обществе два 1-агента при 49 агентах во фракции $G_4$ (рис. 14).

Рис. 18. Клика $G_4$ – "диктатор". Ее СПК неотрицателен, но при $\mu \leq -7$ общество разоряется.

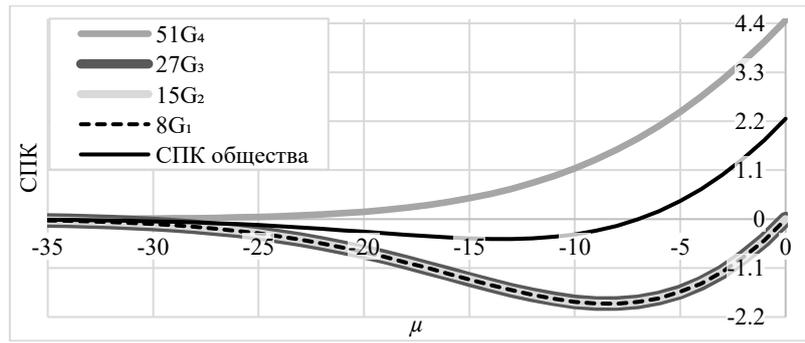

СПК агентов в обществе 101:$\{8G_1; 15G_2; 27G_3; 51G_4\}$.

### 3.4. Сравнение рассмотренных обществ

СПК агентов в некоторых из рассмотренных выше обществ представлен на рис. 19. Общества перечислены ниже. В каждом из них 101 агент; этот общий параметр в описаниях опущен. Результаты соответствуют $\mu = -14$.

0. 101E
1. $\{8S; 93E\}$
2. $\{8(0{,}945S+0{,}055G); 93E\}$
3. $\{8(0{,}934S+0{,}066G); 93E\}$
4. $\{8G; 93E\}$
5. $\{8G_1; 15(0{,}972S+0{,}028G_2); 78E\}$
6. $\{8G_1; 15(0{,}914S+0{,}086G_2); 78E\}$
7. $\{8G_1; 15G_2; 78E\}$
8. $\{8G_1; 15G_2; 27(0{,}9605S+0{,}0395G_3); 51E\}$
9. $\{8G_1; 15G_2; 27(0{,}9S+0{,}1G_3); 51E\}$
10. $\{8G_1; 15G_2; 27G_3; 51E\}$
11. $\{8G_1; 15G_2; 27G_3; 25(0{,}9S+0{,}1G_4); 26(0{,}94S+0{,}06G_5\}$
12. $\{8G_1; 15G_2; 27G_3; 51(0{,}9S+0{,}1G_4)\}$
13. $\{8G_1; 15G_2; 27G_3; 51G_4\}$

Рис. 19. Эволюция общества: при образовании каждой ответственной элиты разорение фракций прекращается. Превращение ее в клику приводит к разорению всех фракций, кроме нее.

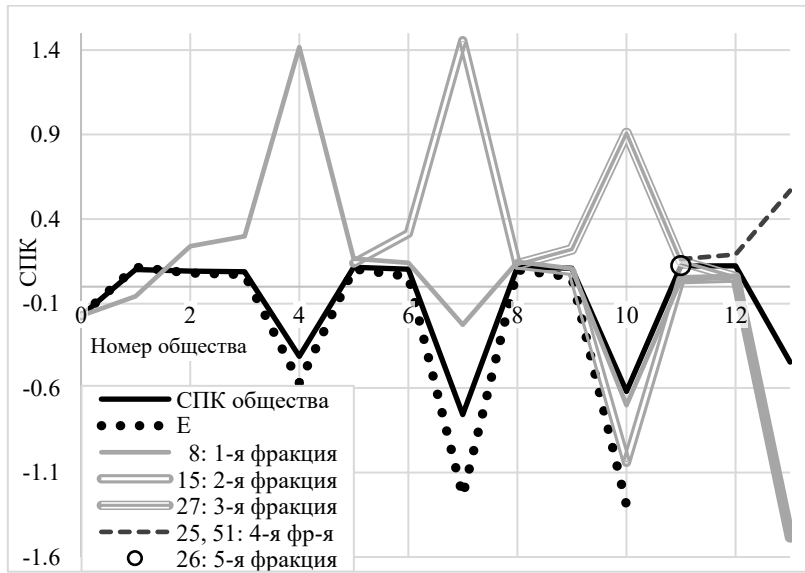

Серия обществ: 101 агент; генератор $N(-14, 80)$. Последовательно образуются и затем становятся кликами четыре ответственные элиты (из 8, 15, 27, 51 агентов).

На горизонтальной оси диаграммы отмечены номера обществ. Исходное общество состоит из индивидуалистов и находится в "яме ущерба": агенты теряют капитал в результате решений большинства. В четырех обществах (4, 7, 10 и 13) СПК всего общества существенно ниже, чем в исходном. Это свидетельство неоднозначности кооперации: она очень выгодна для кооперирующейся клики (вследствие ее влияния на принимаемые решения), но совокупно в еще большей мере невыгодна для остальных. Тем самым кооперация в форме клик не оставляет иной альтернативы, кроме адекватных форм кооперации, агентам, ею еще не охваченным. Варианты этой кооперации: присоединение к имеющимся кликам (если они не блокируют его), формирование новых клик, комбинированные стратегии. Выгодность этих вариантов для участников и совокупно для общества определяется вероятностями реализации всевозможных минимальных решающих коалиций; эти вероятности, в свою очередь, зависят от параметров среды.

Для остальных (кроме исходного и четырех перечисленных) обществ СПК всего общества положителен благодаря присутствию ответственной элиты, размер которой в данном случае – несколько менее удвоенного размера наибольшей клики. Обеспечение положительного СПК общества оказывается совместимым для ответственных элит с собственным умеренным лидерством по доходам (общества 3, 6, 9, 11 и 12).

Отметим общество 11: 1-агенты общества 10 образуют в нем две фракции (25 и 26 агентов), комбинированные стратегии которых имеют разные коэффициенты $\alpha$. Затем в обществе 12 эти фракции соединяются в одну. Сравнение обществ 11 и 12 показывает, что ответственная элита может разделяться на автономные фракции без нарушения определяющих ее свойств.

## 4. Выводы

В предположениях модели ViSE в работе рассмотрены трансформации исходного общества, где решения большинства систематически приводили к разорению агентов в силу парадокса "ямы ущерба".

1. При появлении и росте в таком обществе новой просоциальной фракции в выигрыше оказываются 1-агенты и общество в целом. Они мало теряют, если эта фракция становится ответственной элитой с невысокой долей групповой составляющей комбинированной стратегии голосования. Сравнительно немногочисленная фракция этого типа стабилизирует общество, устраняя парадокс "ямы ущерба". Минимальная доля такой элиты в обществе снижается при увеличении его размера.
2. Для обеспечения ответственной элите дохода, несколько превышающего средний доход по обществу, достаточно невысокой доли групповой составляющей ее целевой функции.
3. Если ответственная элита радикально увеличивает вес групповой составляющей своей целевой функции, ее доходы резко возрастают, в то время как доходы общества существенно снижаются. В этом случае 1-агенты выигрывают от присоединения к элите, но элите выгодно поддерживать свою численность на низком уровне, что обеспечивает ее членам максимальный доход (см. (Tsodikova, 2020, рис. 6)).
4. Общество и индивидуалисты оказываются в убытке, если ответственная элита становится кликой, т.е. фракцией с групповой целевой функцией. Клика чаще всего проигрывает при появлении большей клики, однако, в силу закона больших чисел, меньшая клика опережает бо́льшую, если обе клики являются ключевыми в одних и тех же (по составу остальных участников) решающих коалициях.
5. Если в противовес сформировавшейся элите-клике, разоряющей общество, оно формирует ответственную элиту, превосходящую первую по численности немного менее, чем вдвое, то последней удается остановить общее снижение капитала, поднять доходы своих членов на уровень, несколько превышающий средний по обществу, и значительно снизить доходы клики. При достаточной численности ее члены *лидируют* (без существенного отрыва) в обществе по доходам, поддерживая на положительном уровне и общий средний доход.

6. Если ответственная элита, конкурировавшая с кликой, сама становится кликой, то конкуренция двух сравнимых клик предпочтительнее для общества, чем присутствие одной клики (см. (Чеботарев и др., 2008, рис. 2)). Подобную итеративную социальную динамику можно продолжить: возможность появления ответственной элиты, помогающей обществу и не жертвующей собой, сохраняется на нескольких этапах. Такую динамику можно рассматривать как противостояние ответственных элит эгоистическим кликам.
7. При росте фракции с комбинированной стратегией, близкой к просоциальной, разнонаправленно меняются вероятности образования минимальных решающих коалиций, включающих клики, поэтому средние приросты капитала последних могут меняться немонотонно.

Общий вывод: наличие ответственной элиты в обществе выгодно как ему, так и этой страте, однако она сталкивается с искушением пренебречь ответственностью, сосредоточившись на собственных интересах. Негативные следствия этого могут быть нивелированы, пока возможно формирование элиты большей численности, готовой соблюдать взаимовыгодный общественный контракт.

Изученная динамика выявляет двойственные последствия кооперации. В модели ViSE чистая кооперация представлена кликами, которые обеспечивают высокий доход своим членам, но, как правило, разоряют 1-агентов и членов малых клик, снижая их влияние. В свою очередь это стимулирует последних к эффективной кооперации. Поскольку модель служит упрощенным образом реальности, есть основания (подкрепленные наблюдениями) считать, что описанные механизмы и закономерности могут проявляться и в последней.

## Литература